\documentclass[12pt]{article}
\usepackage{graphics}
\begin{document}
\begin{titlepage}
\begin{flushright}
OUTP-97-P-\\
\end{flushright}

\begin{center}
{\large \bf Effect of retardation on dynamical mass
generation in two-dimensional QED at finite temperature}\\
 \  \\
by\\
\ \\
D.J. Lee\\
\  \\
Department of Physics\\
Theoretical Physics\\
1 Keble Road\\
Oxford   OX1 3NP\\
U.K.\\
\vspace{.05in}
{\bf Abstract}
\vspace{.05in}
\end{center}
{\small The effect of retardation on dynamical mass 
generation in $QED$ in two space dimensions at 
finite temperature is studied, in the imaginary time 
formalism. The photon polarization tensor is evaluated 
to leading order in $1/N$ (where $N$ is the number 
of flavours), and simple approximate closed form 
expressions are found for the fully retarded
longitudinal and transverse
propagators, which have the correct $T \rightarrow 0$ limit.
The resulting Schwinger-Dyson equation for the fermion 
mass (at order $1/N$) has an infrared divergence associated 
with the contribution of the tranverse photon propagator;
only the longitudinal contribution is retained, as in 
earlier treatments. For solutions in which the mass is a 
constant, it is found that retardation reduces the value of 
the parameter $r$ (the ratio of twice the mass to the 
critical temperature) from about 10 to about 6, in agreement 
with a similar calculation in the real time formalism.
The gap equation is then solved allowing the mass to depend 
on frequency (but not momentum), thus extending the study 
of retardation to the  variable mass  case for the first 
time.Solutions for $T \neq 0$ are obtained which join 
on smoothly to the correct $T=0$ solution. It is found that 
there is a critical number of flavours, $N_c$,above which 
no mass is generated. The phase boundary in the $N-T$ plane 
is calculated, and agrees qualitatively with that found in 
other variable mass (but non-retarded) calculations. The 
$r$ value remains close to 6.Possibilities for including 
the transverse photon propagator are discussed. }
\vspace{0.05in}
\begin{flushleft}
June 1997\\
  
\end{flushleft}
\end{titlepage}

\begin{center}
\bf{1.Introduction}
\end{center}

The study of quantum electrodynamics in (2+1) dimensions is of considerable interest, due to its possible relevance to long-wave length models of 2-D condensed matter systems, particularly those which 
might apply to high $T_c$ superconductors $[13]$, $[14]$. $QED_3$ may also give us insight into the phenomenon of dynamical symmety breaking in
 theories such as $QCD$. At zero temperature,
a considerable amount of work has been  
 done $[1]-[8]$. In Appelquist et al $[2]$ it was shown that, using a $1/N$ expansion in the Schwinger-Dyson (S-D) equations, there existed a value $N_c=32/\pi^2$ above which no fermion mass was dynamically generated, $N$ denoting the number of fermion flavours in the theory.This was found to be qualitatively 
still true to order $1/N^2$, $N_c$ changing by a factor of 4/3.
$[5]$. In Pennington and collaborators $[3]$, by contrast, no $N_c$ was found; instead, the dynamically generated mass fell exponentially with increasing $N$. This work adopted a more general non-perturbative approach to the S-D equations. On the other hand, an alternative non-perturbative study by Atkinson et al $[6]$ claimed that there was indeed a critical number of flavours.More recently,
Maris $[7]$ has confirmed the existence of an $N_c$ with a value of 
about 3.3,more or less independent of the choice of vertex ansatz, by 
considering the coupled S-D equations for the photon and fermion 
propagators; and Kondo $[8]$ has studied the problem using the non-local
gauge technique.\\

It is important to extend this type of analysis to finite temperature,because of its possible 
  relevance to physical applications. Here the significant
parameters are  $T_c$ and $r$, where $r$ is the ratio of twice the zero temperature fermion
mass to  the critical temperature, $T_c$.$T_c$ is the temperature above which chiral symmetry is restored, and at which there is a phase transition from the superconducting phase to the normal phase in the model discussed in $[13]$. In Dorey and Mavromatos $[9]$ a calculation was done using the S-D equations,
with the Matsubara finite-temperature formalism,
in which the fermion mass $\Sigma $ was taken to be constant, and only
the $\Delta_{00}$ component of the photon propagator was used in the 
instantaneous approximation-that is, all frequency dependence in the photon propagator was neglected. In this calculation  the authors found a value of $r \simeq 10$. If we identify the fermion mass with the order parameter in B.C.S- like theories, this value is much larger than a typical B.C.S value, which is $r \simeq 3.5$. In Aitchison et al $[10]$ the constant mass approximation was relaxed, and instead a momentum -dependent solution $\Sigma(T,{\bf{p}})$
 was calculated. However once again only $\Delta_{00}$ in the instantaneous approximation was retained, and it was found that $r \simeq 10$. The $r$ value was therefore insensitive to this refinement in the calulation, although the values of $T_ck_B$ and $\Sigma(T=0,{\bf{p}}=0)$ had changed considerably. It was shown in Aitchison and Klein $[11]$ that this value of $r$  also survives the inclusion of a form of 
 wavefunction renormalization. In both the calculations of $[10]$ and $[11]$ the critical value of $N$ was $N_c \sim 2$. It is important to stress that  no such $N_c$ existed in the constant $\Sigma$ case $[9]$.\\

One obvious problem with the instantaneous approximation in the Matsubara 
formalism, is that it cannot reproduce the well-studied zero-temperature
limit. This is because, as $T \rightarrow 0$, all frequency components 
should be included, while the instaneous approximation retains only the 
$n=0$ one. To avoid dealing with a large number of frequency components,
the first calculation to include retardation in the S-D equations for $QED_3$
at finite temperature $[12]$ used a real-time formalism, in which the correct 
$T \rightarrow 0$ was ensured. Making the constant mass approximation, it was 
found that $r$ was significantly reduced to a value of about 6, from the 
non-retarded value of about 10.\\

The calculations of $[12]$, however, did not retain the exact expressions for the 
longitudinal and transverse photon self-energies,because of their awkward 
behaviour near zero three-momentum ( the amplitudes are non-analytic at the 
origin, at finite temperature). Instead, a variety of simpler ``average'' 
self-energies were used in order to simplify the calculation and to be able to
compare the results more easily with those of $[9]$. In the imaginary time 
formalism,this difficulty concerning the non-analytic behaviour near zero 
momentum does not arise, and the exact self-enegies can be employed. Our first 
aim in this paper, therefore, is to include retardation effects, in the constant mass approximation, using the imaginary time formalism. In Section 2 we restate
for convenience the results given in
$[13]$ for $\Pi_{\mu \nu}$, the photon  
polarization tensor to leading order in $1/N$. These results involve certain  integrals which we first evaluate numerically, and then
find simple approximate closed form expressions which retain the correct $T \rightarrow 0$ limit. In section 3 we shall then extend the (constant mass) work done in $[9]$ by using a fully retarded longitudinal
 propagator. We  compare the results with $[12]$, and find rather close 
agreement. \\

 In section 4 we shall go further and allow the fermion mass to 
depend on the (discrete) frquency,still retaining the fully retarded 
longitudinal self-energy. We believe this is the first time that retardation 
has been introduced into a ``non-constant-mass'' S-D calculation
at finite temperature. 
Although we cannot compare our results directly with $[10]$, there are interesting features which remain the same, namely $N_c \sim 2$ and a $p_0/\alpha$ behaviour similar to that of the $|\bf{p}|/\alpha$ behaviour of the momentum dependent solution. Our 
values of $\Sigma(p_0=0)$ are found to be in rough agreement with those of $\Sigma({\bf{p}} = 0) $  in $[10]$. It would,
of course,be interesting to include a dependence  of 
$\Sigma$ on momentum as well as on frequency, but that is a much harder 
problem in the retarded than in the instantaneous case.\\

 In section 5 we shall discuss ways of introducing a transverse contribution into our mass-gap equation, which we neglected due to a logarithmic i-r divergence in the zeroth mode. We shall also discuss other possible extensions to our calculation, and we conclude by looking at the plausiblity of our calculations in the context of $QED_3$ as  a model of superconductivity.

\begin{center}
\bf{2.An approximate form for the full photon propagator 
to leading order in $1/N$}
\end{center}

The Lagrangian of massless $QED_3$ with $N$ flavours is
\begin{equation} L=-1/4f_{\mu \nu}f^{\mu \nu} + \sum_i \bar{\psi_i}( i \partial\!\!\!/ 
-ea \!\!\!/ )\psi_i \end{equation}
 where $a_\mu$ is the vector potential and $i=1,2...N$.We have also chosen a reducible representation of four-dimensional matrices for the Dirac algebra. Due to the choice of representation eqn (1) has continuous chiral symmetry as discussed in $[2]$.\\

Following $[13]$ we calculate the full photon propagator in this theory to leading order in $1/N$ . We then look for simple closed form expressions to approximate those integrals in the calculation which cannot be evaluated analytically. In these calculations we shall be working in the Landau gauge, in Euclidean space.
We first note that the most general photon propagator in the Landau gauge must take the form:
\begin{equation} \Delta_{\mu \nu} = {A_{\mu \nu} \over p^2+\Pi_A(p)}+{B_{\mu \nu} \over p^2+\Pi_B(p)} \end{equation}
where $A_{\mu \nu}$ is the longitudinal projection operator and $B_{\mu \nu}$ is the transverse projection operator. To leading order in $1/N$,$\Pi_A$ and $\Pi_B$
are the contributions from the loop diagrams shown in fig.1 and are related to the polarization tensor, $\Pi_{\mu \nu}$ ,at order $1/N$.
  The projection operators take the following forms:
\begin{eqnarray}
 A_{\mu \nu} &=& \left( \delta_{\mu 0} - {p_{\mu} p_{0} \over p^2} \right){p^2 \over \bf{p^2}} \left( \delta_{0 \nu } - {p_{0} p_{\nu} \over p^2}  \right)
  \nonumber \\
\mbox{and} \;\; &B_{\mu \nu}& = \; \; \; \delta_{\mu i} \left( \delta_{ij}-{p_i,p_j \over \bf{p^2}} \right)
 \delta_{j \nu} .
\end{eqnarray}
By using the properties of the projection operators it is easy to show that the inverse propagator must be:
\begin{equation} 
\Delta^{-1}_{\mu \nu}(p) = (p^2 + \Pi_A(p))A_{\mu \nu}(p)+(p^2 + \Pi_B(p))B_{\mu \nu}(p).
\end{equation}
>From this expression it is easy to relate $\Pi_{\mu \nu}$ to $\Pi_A$ and $\Pi_B$
by summing the diagrams in fig.1 :
\begin{equation}
\Pi_{\mu \nu}=\Pi_A A_{\mu \nu} + \Pi_B B_{\mu \nu}.
\end{equation}
Then using the explicit forms of $A_{\mu \nu}$ and $B_{\mu \nu}$ in (3) with the additional requirement that $p_{\mu} \Pi^{\mu \nu} =0$ one obtains the following forms of $\Pi_A$ and $\Pi_B$:
\begin{eqnarray}
   \Pi_A &=& \Pi_{00} {p^2 \over {\bf p^2}} \nonumber \\
    \Pi_B &=& \Pi_{ii}-\Pi_{00}{p_0^2 \over p^2}.
\end{eqnarray}
We now proceed with the calculation of $\Pi_{\mu \nu}$ at finite temperature using the Matsubara formalism. We work in imaginary time in which both fermonic and bosonic frequencies are discrete. Fermionic frequencies have the form $p_{0f}={2\pi \over \beta}(m+1/2)$ and bosonic frequencies
 have the form $p_{0b}={2\pi \over \beta}m$, where $m$ is an integer. From now on we shall denote the modulus of 
bosonic 3-momenta by $p_b$ and that of
 fermionic 3-momenta by $p_f$.In our calculation we need only consider the elements $\Pi_{00}$ and $\Pi_{ij}$. $\Pi_{00}$ and $\Pi_{ij}$ are calculated to be $[13]$:
\begin{eqnarray}
 \Pi_{00} &=& \Pi_3 - {p_{0b}^2 \over p_b^2}\Pi_1 - \Pi_2 \nonumber \\
 \Pi_{ij} &=& \Pi_1 - \left( \delta_{ij} - {p_{i}p_{j} \over p_b^2 }\right) + \Pi_2 \delta_{ij}
\end{eqnarray}
where
\begin{eqnarray}
\Pi_1(p_b,\beta,m) &=& {\alpha p_b \over 2\pi}\int_{0}^{1}dx \sqrt{x(1-x)}{ \sinh (p_b \beta \sqrt{x(1-x)}) \over D_m(x,p_b,\beta)} \nonumber \\
\Pi_2(p_b,\beta,m) &=& {\alpha m \over 2 \beta} \int_{0}^{1}dx (1-2x){ \sin (2xm \pi ) \over D_m(x,p_b,\beta)} \\
\Pi_3(p_b,\beta,m) &=& {\alpha \over \pi \beta} \int_0^1 dx \ln(4D_m(x,p_b,\beta)) 
\nonumber
\end{eqnarray}
and
\begin{equation}
D_m(x,p_b,\beta)=\cosh^2 ( \beta p_b \sqrt{x(1-x)} / 2 )
-\sin^2 (xm \pi).
\end{equation}\\
$m$ denotes explict frequency dependence in the above expressions.\\
Using the result 
\begin{equation}
\Pi_3=\Pi_1 + {p_b^2 \over p_{0b}^2}\Pi_2
\end{equation}
we find that $\Pi_{A}=\Pi_3$,$\Pi_{B}=\Pi_{1}+\Pi_{2}$.
Alternatively one can obtain the above expressions for $\Pi_A$ and $\Pi_B$ by neglecting $\Pi_2$ and setting $\Pi_1=\Pi_3$, which is found to be a good approximation for $\Pi_{ij}$ and $\Pi_{00}$ when $m \neq 0$. For $m=0$ one can find these expressions without again using eqn (10), by noticing 
that $\Pi_2(m=0)=0$ and $p_{0b}=0$.\\

In our treatment of $\Pi_1$, $\Pi_2$ and $\Pi_3$ it will be convenient to consider the $m=0$ mode separately. For $m=0$ we need only consider $\Pi_1$ and $\Pi_3$,since  $\Pi_2(m=0)=0$. For $\Pi_3(m=0)$, which we shall denote by $\Pi^0_3$, an accurate closed form expression has been given in $[10]$, namely
\begin{equation}
\Pi_3^0 = 1/8 \left( {\alpha \over \beta} \right) \left[ | {\bf{p}} | \beta + { 16 \ln 2 \over \pi} \exp \left( -{\pi \over 16 \ln 2 } | {\bf{p}} | \beta \right) \right].
\end{equation}
We are able to find a similar expression for $\Pi_1^0 =\Pi_1(m=0)$ by noting that
\begin{equation}
\Pi_1^0 = ( {|\bf{p}|}) {\partial \over \partial {|\bf{p}|} \beta } \Pi_3^0
\end{equation}
which gives, on combining eqn (11) and eqn (12),
\begin{equation}
\Pi_1^0 =  {\alpha \over \beta} \left( {{|\bf{p}}| \beta \over 8} \right) \left[ 1 - \exp \left( -{\pi \over 16 \ln 2} | {\bf{p}} | \beta \right) \right].
\end{equation}
We now evaluate the integrals $\Pi_1(p_b,\beta,m)$, $\Pi_2(p_b,\beta,m)$ and $\Pi_3(p_b,\beta,m)$ for $m \neq 0$. We show the numerical results of these as 
functions 
 of $p_b \beta$ and $m$ in figs 2a, 2b and 2c. We see that both $\Pi_1$ and $\Pi_3$ vary little with the explicit
 frequency index $m$. Also we are able to show that $0 < \Pi_2 < \left( {2 \ln 2 \over \pi } \right) \left( {\alpha \over \beta} \right)$.
$\Pi_1$ and $\Pi_3$ are well approximated by
\begin{equation}
\Pi_1 = \Pi_3 \simeq 1/8 \left( \alpha \over \beta \right) \left( p_b \beta \right)
\end{equation}
for $m \neq 0$ as figs 3a and 3b show. Since $\Pi_2$ is bounded by ${2\ln2 \over \pi} \left( \alpha \over \beta \right)$ it is always smaller than $\Pi_1$ for $m \neq 0$, since $\Pi_1 \geq \left( {2\pi \over 8} \right) \left( {\alpha \over \beta}
\right) $ ,so we shall neglect it in $\Pi_B$.\\

We now have two expressions for the full photon propagator:
\begin{eqnarray}
\Delta_{\mu \nu} &\simeq& { A_{\mu \nu} \over |{\bf{p}}|^2 + \Pi_3^0 (|{\bf{p}}|) }
+{ B_{\mu \nu} \over |{\bf{p}}|^2 + \Pi_1^0 (|\bf{p}|) },
\mbox{for }  p_{0b}=0 \nonumber \\
\nonumber \\
\Delta_{\mu \nu} &\simeq& { A_{\mu \nu} \over p_b^2 + \alpha p_b/8 }
+{ B_{\mu \nu} \over p^2 + \alpha p_b/8 },
\mbox{for }   p_{0b} \neq 0  . 
\end{eqnarray}
The structure of our propagator is consistent with the zero temperature result, to order $1/N$: $\Pi_A=\Pi_B=\alpha p/8$. One should note that although in the zero temperature limit the theory is Lorenz invariant, Lorenz invariance is clearly broken at $T \neq 0$, as (15) shows. This is, of course, because a preferred frame of reference is provided by the heat bath.\\

\begin{center}
 \bf{3 The Schwinger-Dyson equation
and its solution for constant $\Sigma$}
\end{center}
 
The full Schwinger-Dyson equation for the fermion propagator at non-zero temperature $k_BT=1/\beta$ is given by
\begin{equation}
S^{-1}_F(p_f)=S^{(0)-1}_F(p_f) - {e \over \beta} \sum_{n = - \infty }^{ \infty }
\int {d^2k \over (2 \pi)^2 } \gamma^{\nu} 
 S_F(k_f) \Delta_{\mu \nu}(k_f-p_f) \Gamma^{\mu}(k_f-p_f,k_f).
\end{equation}
We now truncate eqn (16) by working to leading order in $1/N$, in which case $\Gamma^{\nu}$ is replaced by its bare value $e \gamma^{\nu}$ and we use our form for $\Delta_{\mu \nu}$, the full photon propagator, given in (15). As before  $[9]$ we neglect wave function renormalization; we comment further on this in the next section. On taking the trace of eqn (16) we get the following equation for the mass-gap function $\Sigma$ \\
\begin{equation}
\Sigma(p_f) = {\alpha \over N \beta} \sum^{\infty}_{n=- \infty} \int {d^2k \over ( 2 \pi)^2 } \Delta_{\mu \mu}(k_f-p_f) {\Sigma(k_f) \over k_f^2 +\Sigma^2(k_f)}
\end{equation}
where
\begin{eqnarray}
\Delta_{\mu \mu}(q_b=k_f-p_f) &=& {1 \over |{\bf{q}}|^2+\Pi_1^0(|{\bf{q}}|) }+{1 \over |{\bf{q}}|^2+\Pi_3^0(|{\bf{q}}|) } ,
 \mbox{for }    q_{0b}=0 \nonumber \\
\nonumber \\
\Delta_{\mu \mu}(q_b) &=& {2 \over q_b^2+\alpha q_b/8 } , 
\mbox{for }    q_{0b} \neq 0 
\end{eqnarray}
and $\alpha=Ne^2$. Now we make the approximation that $\Sigma$ is frequency as well as momentum independent. We can then remove a factor of $\Sigma$ from each side of the equation, and write the argument of  $\Delta_{\mu \mu}$
 in (17) as $( {2 n \pi \over \beta }, {\bf k} )$. We now do the angular integration, which is trivial. Rearranging terms  gives 
\begin{equation}
1={a \over 2N \pi} [S_L(a,s)+S_T(a,s)].
\end{equation}
$S_T$ is the transverse contribution which is expressed as:
\begin{eqnarray}
S_T(a,s) &=& \int^{\infty}_{0} x \; dx   { 1 \over x^2 + \beta^2 \Pi_1^0 }
\; { 1 \over x^2 + \pi^2 + a^2s^2 } \nonumber \\
\nonumber \\
&\; +& \sum_{m=1}^{ \infty } \left( { 1 \over x^2 + (2 \pi m)^2 +
 0.125 a (x^2 + (2 \pi m )^2)^{1/2} } \right)  \\
\nonumber \\
&\; \times&  \left( { 1 \over x^2 + (2 \pi(m+1/2))^2 +a^2s^2 } + { 1 \over x^2 + (2\pi (m-1/2))^2 +a^2s^2 } \right) \nonumber \\
\nonumber
\end{eqnarray}
where $x = \beta | \bf{k} |$,$a=\alpha \beta$, and $s=\Sigma/\alpha$. $S_L$ is the 
longitudinal contribution and is exactly the same, except for $\Pi_1^0$ being replaced by $\Pi_3^0$. It is now important to notice that the integral for the 0th transverse mode is logarithmically i-r divergent. This can be seen by observing that as $\beta|{\bf{k}}| \rightarrow 0$, $\Pi_1^0 \rightarrow  \left( { \alpha \over \beta} \right)  \left( { \pi \over 128 \ln 2 } \right) |\bf{k}|^2\beta^2 $. From now on, we
  shall  retain only the longitudinal mode; in our Conclusion we shall discuss ways of including the transverse mode in our calculations. \\

So we consider the following equation
\begin{equation}
1 = {a \over 2 \pi N}S_L(a,s).
\end{equation}
The integral for the $m \neq 0$ modes can be done analytically. We introduce functions of the form:
\begin{equation}
I(d,a,c) = \int_{0}^{\infty} {x \; dx \over (x^2+d^2) +a(x^2+d^2)^{1/2}}{1 \over x^2 +c^2}
\end{equation}
Doing the integral on the R.H.S of eqn (22) gives us the following closed form expressions for $I(d,a,c)$.
\begin{eqnarray}
I(d,a,c) &=& {1 \over 2(a^2+c^2-d^2)} \ln \left( {c^2 \over (d+a)^2} \right)  \\
&+& 
{a \over (c^2-d^2)^{1/2}(a^2+c^2-d^2) } \arctan \left( (c^2-d^2)^{1/2} \over d \right) .\nonumber
\end{eqnarray}
With the use of these functions we are able to write $S_L(a,s)$ as
\begin{eqnarray}
S_L(a,s) &=& \sum_{m=1}^{\infty} I( 2 \pi m ,0.125a,(a^2s^2+(2 \pi (m+1/2))^2)^{1/2}) \nonumber \\
&+&I ( 2 \pi m ,0.125a,(a^2s^2+(2 \pi (m-1/2))^2)^{1/2})  \\
&+& \int_0^{\infty}{x \; dx \over (x^2+ \Pi_3^0(x)\beta^2)(x^2+\pi^2+a^2s^2) } .\nonumber
\end{eqnarray}
We now solve eqn (21) numerically. We do this by fixing $a$ and $N$ and varying $s=\Sigma/\alpha$ until the R.H.S is equal to unity.
We note the value of $s$ for which eqn (21) is satisfied for those values of $N$ and $a$. We then choose new values of $a$ and $N$ and repeat the process. If eqn (21) is not satisfied for any $s$ at the values of $a$ and $N$ chosen, the  only solution to eqn (17) for constant mass is the trivial one $s=0$. It is 
important to note that, in these calculations, the infinite sums are 
evaluated numerically using Mathematica: the convergence is very slow if 
an attempt is made to truncate them at a (large but) finite number of 
terms.\\

In our analysis we also require the zero temperature limit of eqn (21) which is found to be
\begin{equation}
1 = {1 \over (2 \pi)^2 N} \int_{0}^{\infty} {x \; dx \over (x^2+s^2)(x+0.125) }
\end{equation}
where $x=|\bf{k}|/\alpha$.
On doing the integration on the R.H.S one obtains the following equation
\begin{eqnarray}
(2 \pi)^2 N/2 = \left( 1-{(0.125)^2 \over (0.125)^2+s^2} \right) {\pi \over 2s}
\nonumber \\
-{(0.125) \over (0.125)^2+s^2}\ln \left( {s \over 0.125 } \right).
\end{eqnarray}
Now if $s << 0.125$ we get the approximate solution $ s \simeq 0.25 \exp (-
{\pi}^2 N/4)$, in agreement with $[1]$,remembering that we are including 
only the longitudinal self-energy, which contributes one half of the total 
contribution at zero temperature, so that our ``$2N$'' corresponds to 
Pisarski's ``$N$''.In fact, we have found that a more accurate fit to the 
exact numerical solution of (25) is provided by
\begin{equation}
s \simeq 0.25 \exp (-0.96 \pi N).
\end{equation} \\
With the zero temperature result and our data for $T \neq 0$, we are able to plot solutions of (21) as a function of 
 $1/a=k_BT/\alpha$ for different values of N, as shown in fig.4. It is important to see that unlike previous calculations our zero temperature results join on smoothly to our finite temperature results. This is because we have used a retarded form of the 
photon propagator with the correct $T \longrightarrow 0$ limit, and not the instantaneous form used in $[9]$. Another thing to note is that the shapes of the solutions as functions of $k_BT/ \alpha$ are markedly different from those of $[9]$, $[10]$ and $[11]$; they seem more to resemble the
shape of the
 B.C.S constant mass-gap solution.
We also plot  $T_ck_B / \alpha$ against $N$,  in fig.5. $T_ck_B$ falls
 off exponentially and takes the following form, to very high accuracy:
\begin{equation}
(T_ck_B)/ \alpha \simeq 0.081 \exp (-0.96 \pi N).
\end{equation}
>From eqn (27) and eqn (28) we can calculate $r$,where $r={2 \Sigma(T=0) \over
T_ck_B}$;we find $r=6.17$.\\

In this calculation there is therefore no $N_c$. The reason for this can be seen as 
follows.We first note that increasing $s$ and decreasing $a$ (increasing $T$) causes the R.H.S of eqn (21),${a \over 2 \pi N}S_L$, to get smaller. The 
condition determining $N_c$ 
is that $\lim_{a \rightarrow \infty}(aS_L) < 2\pi N_c$ for all $s$,since then eqn (21) can no longer be satisfied, and the only solution to the gap equation is $s=0$. Since $S_L$ is a monotonically decreasing function of $s$ we need only show  that $\lim_{a \rightarrow \infty}(aS_L(s=0))$ is finite to establish the existence of $N_c$. But  one  easily sees from eqn(26) that $\lim_{a \rightarrow \infty}(aS_L(s=0))$
is actually infinite, and so $N_c$ does not exist.\\

In $[9]$ the instantaneous approximation was used with the constant gap approximation, giving a higher value $r$, $r \simeq 10$. There is good agreement  between our results and those of $[12]$,  in which retardation was included 
in a real time formalism. For example,in $[12]$, using the 
``best average self-energy'' ($\Pi^{R1}$ in the notation of $[12]$),$s$ had 
the value 0.01205 for $N=2$, corresponding to a value of $s=0.1225$ from (27) 
with $N=1$;and $(k_B T_c)/ \alpha $ was 0.0376, corresponding to 0.0397 from (28). These values lead  to $r=6.41$ in $[12]$ as compared to our value of 6.17. The only other difference between our results and those of $[12]$-not a large one-is that we see very little change of $r$ with $N$,whereas $[12]$ found a small
variation with $N$.\\

 In the following section we shall look at solutions which are frequency dependent and contain the retarded photon propagator calculated in section 2.\\

\begin{center}
 \bf{4 The fequency dependent gap equation}
\end{center}
 
In this section we depart from the constant gap approximation, and attempt to solve the retarded
 Schwinger-Dyson equation (17) allowing the gap function to depend on frequency only. Frequency dependence can be introduced relatively  easily into the equation, unlike momentum dependence which leads to a highly non-trivial angular integral of the kernel of our equation, which is unlikely to have an analytic result. The resulting computational complexity has deterred us from including momentum dependence.This is unfortunate, for it would be interesting to compare results calculated with a full retarded 

momentum dependent S-D equation with the results of $[10]$. While this is true, introducing frequency dependence will give us new insight into the nature of the gap, and as in $[10]$ we shall be able to calculate a value of $N_c$. In the spirit of $[10]$, 

we again neglect wavefunction renormalization. It has been pointed out in $[3]$ that wavefunction renormalization should be introduced, but it has been found in $[11]$ that for a momentum dependent S-D gap equation, there still exists an $N_c$ and $r$ is 
little affected.\\

We start our study of the frequency dependent 
solution at eqns (17) and (18). We shall restrict 
ourselves again only to 
the longitudinal mode; this is because the i-r 
divergence in the tranverse mode
 is still present, although only in the mode for which 
$q_{0b }=k_{0f}-p_{0f}=0$. We see
 that because $\Sigma$ is now frequency dependent we can no longer 
remove a factor of $\Sigma$ from each side of eqn (17),
 and $\Delta_{\mu \mu}$ is now dependent on  
$p_{0f}={2\pi \over \beta}(n+1/2)$ . 
But the angular integration is again trivial. 
On rearrangement we get:
\begin{equation}
s(n)={a \over 2 \pi N}S^n_L(a,s) 
\end{equation}
where as before $s=\Sigma/\alpha$ and $a=\alpha\beta$.
$S^n_L$ is the frequency dependent longitudinal 
contribution, expressed as:
\begin{eqnarray}
S^n_L(a,s(n)) &=& \int_0^{\infty} x \; dx {s(n) \over (x^2+\beta^2\Pi_3^0)(x^2+(2\pi(n+1/2))^2+a^2s^(n))} \nonumber \\
&+& \sum_{m \neq n}^{\infty} {s(m) \over x^2+(2\pi(m-n))^2+0.125a(x^2+(2\pi(m-n))^2)^{1/2}} \nonumber \\
&\times& {1 \over x^2+(2\pi(m+1/2))^2+a^2s^2(m)} 
\end{eqnarray}
where $x=\beta|\bf{k}|$.The first term represents the $m=n$ mode,
 where $q_{0b}=0$. Again it is possible 
to do the $x$-integration. Using the functions, 
$I(d,a,c)$ we are able to express $S^n_L$ as:
\begin{eqnarray}
S^n_L(a,s(n)) &=& \sum_{m \neq n}^{\infty} s(m)I(2\pi |m-n|,0.125a,(a^2s^2(m)+(\pi(2m+1))^2)^{1/2}) \nonumber \\
&+& \int^{\infty}_{0}{x \; dx \over  x^2+\beta^2\Pi_3^0 }{s(n) \over x^2+(2\pi(n+1/2))^2+a^2s^2(n) }
\end{eqnarray}
Eqn  (29) represents an infinite system of non-linear 
coupled integral equations, one for each mode, $m$. 
We can reduce the number of equations that need to 
 be considered by the following particle-antiparticle
 symmetry property,
\begin{equation}
s(m)=s(-m-1) \;\; \mbox{where}\;m \geq 0.
\end{equation}
It is important that the R.H.S of eqn (32) is $s(-m-1)$ 
and not $s(-m)$, since $s(m)$ is a fermonic quantity;
 $p_{0f}$ is related to $m$ by
$p_{0f}={2\pi \over \beta}(m+1/2)$. 
Therefore $m$ and $-m-1$ represent $p_{0f}$ and $-p_{0f}$ respectively.
 Using eqn (32) it is possible to cast $S^m_L(a,s(n))$ into a final 
form convenient for numerical calculations:
\begin{eqnarray}
S^n_L(a,s) &=& \sum_{m \neq n \geq 0}^{\infty} s(m)[I(2\pi|m-n|,0.125a,(a^2s^2(m)+((\pi(2m+1))^2)^{1/2}) \nonumber \\
\nonumber \\
&+& (I(2\pi|m+n+1|,0.125a,(a^2s^2(m)+((\pi(2m+1))^2)^{1/2})] \nonumber \\
\nonumber \\
&+& \int_0^{\infty}{ x \; dx s(n) \over (x^2+\beta^2\Pi_3^0(x))(x^2+(\pi(2n+1))^2+a^2s^2(n)}  \\
\nonumber \\
&+& s(n)I(2 \pi (2n+1),0.125a,(a^2s^2(n)+((\pi(2n+1))^2)^{1/2}) ,\nonumber \\
\nonumber
\end{eqnarray}
the last term being the $n=-m-1$ mode in eqn (31). Although eqn (33) looks 
cumbersome compared with eqn (31),
 the advantage is now that only positive values of 
 $n$ and $m$  need be considered, thereby 
reducing the number of equations in the system (29).\\

In our numerical calculation we will want to limit the 
number of equations to a finite number, $M$. 
We can think of eqn (29) as a sort of non-linear matrix problem,
 where instead of an infinite matrix we will restrict ourselves 
to an $M \times M$ matrix. The justification for doing this is simple:
we expect our solution to fall off monotonically with $n$, and 
as $n \rightarrow \infty$ we expect $s(n) \rightarrow 0$. The effective limit $M$ on the matrix size is determined by the requirement that $s(n-1=M)$ 
be one tenth the size of $s(0)$,
  which is the maximum value of $s$ 
as a function of $n$. In choosing this size of matrix, the contribution 
from modes for which $n-1 > M$ has a very small effect upon our solution.
Note that whereas in the constant mass case (Section 3) we had to 
include all frequencies in our summations,when the mass is frequency-
dependent a natural cut-off emerges.\\

The method we shall normally adopt (  ``method I'') in solving the
equations (29) will be to start with a trial function
  having the correct 
$n \rightarrow \infty $ limit, namely $s(n) \rightarrow 0$. 
We insert this trial function into eqn(33). 
Then we work out the resultant value of $s(n)$ using eqn (29), and
reinsert the result back into eqn (33). 
After several  such iterations we check if our matrix size is sufficient,
 by seeing if $s(n-1) \leq 0.1s(0)$; 
if it is appreciably larger we increase the size of our matrix.
To do this,
we increase the value of $m$ at which we truncate the sum in eqn (33), 
and we increase  the number of equations in eqn (29) 
we must solve. When we are satisfied with the size of our matrix
 we keep iterating  until s(n) changes by less than $1\%$ over ten 
iterations (or more depending on the rate of convergence 
to the exact solution). It should be appreciated that the 
non-linearity of the system of equations requires 
this further iteration, even after deciding on 
the size of the matrix.\\

If the rate of convergence is very slow, 
which is the case near a phase transition, we use ``method II''. 
In method II we choose two trial functions. 
One of them, $\tilde{s_{\downarrow}}$,
is chosen to decrease in magnitude with each iteration 
using method I; 
the other,$\tilde{s_{\uparrow}}$, 
to increase in magnitude with each iteration. 
We then average the two functions together, $s=0.5(\tilde{s_{\uparrow}}+\tilde{s_{\downarrow}})$. 
Then we use method I on $\tilde{s}$ for ten iterations or more. 
If $\tilde{s}$ increases in magnitude we set $\tilde{s}_{\uparrow}(n)=\tilde{s}(n)$; 
if $\tilde{s}$ decreases in magnitude we set $\tilde{s}_{\downarrow}(n)=\tilde{s}(n)$, and we average again. 
We repeat the process until $\tilde{s_{\uparrow}}$ 
and $\tilde{s_{\downarrow}}$ vary little from each other.\\

An important technical difficulty is that as $T$ gets 
smaller the size of our matrix increases: 
as $T \rightarrow 0$,$M \rightarrow \infty$,and
the larger $M$ is, the slower the calculation of $s(n)$. 
Fortunately most of the interesting behaviour 
lies in a region of temperature where the matrix size, 
$M$,  is manageable. Although we cannot approach close to 
the $T \rightarrow 0$ limit from our finite 
temperature calculations, we can get around this
 difficulty by including an analytic $T \rightarrow 0$ 
limit to eqn (28),
\begin{eqnarray}
s({p_{0f} \over \alpha}) &=& {1 \over (2 \pi)^2 N}\int_0^{\infty} \; dy \; s(y)((I(y-{p_{0f} \over \alpha},1/8,\sqrt{y^2+s^(y)}) \nonumber \\
&+& (I(y+{p_{0f} \over \alpha},1/8,\sqrt{y^2+s^(y)}))
\end{eqnarray}
where $y=k_{0f}/\alpha$. Again we have used 
the antiparticle-particle symmetry property of the gap,
 $s(p_{0f})=s(-p_{0f})$, to reduce the number of equations we 
need consider. Although the range of integration is infinite,
 the kernel of eqn (34) provides an effective cutoff 
$\Lambda$, where $\Lambda < 1$. Therefore we need only
 consider $s({p_{0f} \over \alpha})$ from $s(0)$ to $s(\Lambda)$
 in doing numerical calculations. $\Lambda$ is chosen in the same way as $M$;
 namely at a point where $s(\Lambda)$ is suffiently small
 so that the contributions from $p_{0f}>\Lambda$ are negligible. 
Our $T \neq 0$ results and our $T=0$ results are sufficient to provide
 all the information that we require.\\

We show the results of both our $T=0$ and $T \neq 0$ 
calculations as functions of ${p_{0f} \over \alpha}$ 
for N=1 in  fig.6. One can see that the $T=0$ curve joins
 smoothly on to the $T \neq 0$ solutions. 
One should observe that although we plot the $T \neq 0$ solutions as
 continuous curves, they are actually discrete points 
with $p_{0f}={2 \pi \over \beta}(n+1/2)$. As functions
 of ${p_{0f} \over \alpha}$, these functions fall off rapidly for ${p_{0f} \over \alpha}<0.1$. This behaviour is roughly the same as was seen in  momentum-dependent calculations $[10]$, where for $0.02<{|{\bf{p}}| \over \alpha}<0.1$ the solutions fell rapidly
in ${|\bf{p}| \over \alpha}$. It is this behaviour - the rapid fall in  $s({p_{0f} \over \alpha})$  for ${p_{0f} \over \alpha}<0.1$ - which makes $\Sigma(T=0,p_{0f}=0)$ considerably smaller than $\Sigma(T=0)_{const}$, our result from the previous section, as shown in fig.7.$\Sigma(T=0,p_{0f}=0)$ is roughly
 the same as $\Sigma(T=0,{\bf{p}}=0)$ in $[10]$. This suggests that the constant mass -gap approximation is not a good one, except as an order of magnitude calculation for the quantity $r$. In fig.8a we show the $m=0
$ mode of these solutions for various $N$ as functions of ${k_BT \over \alpha}$, and in fig.8b we show the $m=0$ mode of these solutions for various  ${k_BT \over \alpha}$ as functions of $N$. One notices that the shape of the solutions in fig.8a is the same
 as that of the constant mass-gap solution (fig.4), which we have characterised as B.C.S -like. By looking at fig.8b one can see that as well as a critical temperature, $T_c$, there is also a critical number,$N_c$. $N_c$ is the value of $N$ above which $s=0$ for all $T$ and $p_{0f}$, and this can be found from  fig.8b by locating the point where the gradient of the $T=0$ curve  tends to infinity. We only need consider the $m=0$ mode at $T=0$, since  $s(m)$ is a monotonically decreasing function of both frequency
 and temperature. So if $s(T=0,p_{0f}=0)=0$ this condition must be true for any $T$ and $p_{0f}$. We find that $N_c$ lies in the range  $N_c \sim 1.8-2$.In fig.9 we show a N-T phase diagram, where we have included the value $N_c=1.8$. Here we see that the phase boundary agrees qualitatively with that calculated in [10].\\

 The existence of $N_c$ leads us to the conclusion that any frequency or momentum dependence in the mass-gap equation induces a value of $N_c$. Our calculated value of $N_c$ agrees well with $[10]$, but $[10]$ uses the instantaneous approximation and so has the incorrect  $T \rightarrow 0$ limit. When compared with the correct $T \rightarrow 0 $ limit of the 3-momentum -dependent mass-gap equation of $[2]$, our value for $N_c$ is smaller.(In $[2]$ $N_c$ was calculated to be 3.2). From our results we are able to calculate a table of values for $r$, each at different values of $N$, as shown in table.1. These values are not much different from the value $r \sim 6$ calculated in section 2, but now there is a significant dependence of $r$ on $N$; the values of $r$ are seen to fall with increasing $N$. The insensitivity of the $r$ values to the introduction of a momentum dependence in the S-D equation was seen in $[10]$, although for the instantaneous approximation; again this is roughly seen to be the case with the introduction of frequency dependence as shown by our results.\\

This completes our analysis of the frequency dependent solutions. In the next section we discuss ways of handling the transverse  mode and other ways of extending our calculation.\\

\begin{center}
 \bf{5 Conclusion}
\end{center}
 
As we have seen already in section 4 the constant mass-gap approximation is not a good one to make except possibly for the calculation of $r$, which is little changed by frequency dependence for the range of $N$-values considered here. Also we have seen  that frequency dependence induces an $N_c$. Our result differs from $[2]$, for we have been unable to introduce frequency and momentum dependence into our calculations while preserving a smooth $T \rightarrow 0$ limit,and also because we have no transverse
contribution to the mass-gap equation. The reason why the introduction of momentum dependence is a hard problem at $T \neq 0$ is that three-dimensional Lorenz invarance is lost, since frequency dependence becomes discretized (in the imaginary time formalism). At $T=0$ one can do the angular integration with full 3-momentum dependence, because  one can exploit the three-dimensional symmetry.At $T \neq 0$ this is not possible, due to the preferred (frequency) direction; instead we are faced with an integral in 2-dimensional momentum space, exactly the integral which is discussed at the begining of section 4.\\

Up to now in our analysis we have chosen not to discuss the transverse contribution, and have neglected it from our calculations due to the i-r divergence in its zeroth mode. It is important to stress that the i-r divergence is present only when $T \neq 0$. At $T=0$, the effect of the transverse mode is merely to double the number of flavours, so giving us an $N_c$ in the range  $3.8-4$. To treat the i-r divergence at $T \neq 0$ one needs to regulate the integral in eqn (20). A reasonable way to do this might be
 to introduce massive fermion propagators into our calculation of $\Pi_{\mu \nu}$, the mass of the fermion propagators being calculated self consistently in eqn (19). A major problem with this refinement would be the requirement of self consistency in our calculation of the photon propagator when we have any frequency or momentum dependence in the fermion mass. Nevertheless, it might be worth considering the simpler problem of a constant mass in this sort of calculation; this may not give us  reliable values of $\Sigma(T=0)$ and $k_BT_c$, but the value of $r$ it gives  might  not change very much if frequency or momentum dependence were to be included. To tackle the harder problem of including frequency or momentum dependence in this extension to our work, we may be forced either to consider our calculation of $\Pi_{\mu \nu}$ as a separate integral equation, so that the calculations would involve two integral equations coupled togther, not one; or we may have to use  an ansatz for our expected form of the fermion mass.\\

As well as the above extension to our work involving the transverse contributions, another important step will be to introduce momentum dependence. Although we have stated that this is difficult due to the computational complexity, one may be able to approximate the kernel in such a way as to simplify the problem. One of the major goals in this series of numerical calculations for $T \neq 0$ should be to join on with the results of $[2]$ in a smooth $T \rightarrow 0$ limit, and this necessitates a successful treatment 
of momentum dependence.\\

As to the question of $QED_3$ as a model of superconductivity in the context of our calculations, consider the case  $N=1$.If we crudely double the number of flavours so as to account for the transverse mode, this will effectively count as $N=2$ which is  the value required for $QED_3$ in  a model of high $T_c$ superconductivity $[13]$. We find that $k_BT_c \sim 10^{-3} \alpha$ in agreement with $[10]$, once rescaled to agree with $[2]$ as regards $N_c$.As pointed out in $[10]$ this gives a value of $ \alpha$ of the order of $8 eV$ for $T_c \sim 100^oK$.As stated in $[10]$ this is still much higher than the typical Heisenberg exchange energies, but might be acceptable by suitable rescaling of the fermion lattice operators in the lattice model of $[13]$. It is
interesting to note as a concluding remark that our suggested way to treat the i-r divergence in the zeroth mode of the transverse contribution may effectively reduce $\alpha$ by an order of magnitude or more,so on this basis alone it is a calculation worth considering.\\

\begin{center}
 {\bf Acknowledgements}
\end{center}
D.J.Lee is very grateful to his supervisor I.J.R.Aitchison for useful discussions, and for the editing of the manuscript. He would also like to thank G.Metikas for his diligent proof reading of various formulae given in this paper. The support of a Research Studentship from PPARC(UK) is 
acknowledged.

\begin{center}
 \bf{References}
\end{center}

\noindent $[1]$ R.D Pisarki, $Phys$. $Rev$. {\bf{D29}} (1984) 2423.\\
\\
$[2]$ T.W Appelquist, M.Bowick, D.Karabali and L.C.R Wijewardhana, $Phys$. $Rev$. {\bf{D33}} (1986) 3704.\\
  T.W Appelquist,D.Nash and L.C.R Wijewardhana $Phys$. $Lett$. {\bf{60}} (1988) 2575.\\
\\
$[3]$ M.R Pennington and D.Walsh, $Phys$. $Lett$ {\bf{B253}} (1991) 246.\\
\\
$[4]$ E.Dagotto, A.Kocic and J.B Kogut, $Phys$. $Rev$. $Lett$. {\bf{62}} (1989) 1083 and $Nucl$. $Phys$. {\bf{B334}} (1990) 279.\\
\\
$[5]$ D.Nash,$Phys$. $Rev$. $Lett$. {\bf{62}} (1989) 3024.\\
\\
$[6]$ D.Atkinson, P.W Johnson and P.Maris ,$Phys$. $Rev$. {\bf{D42}} (1990) 602.\\
\\
$[7]$ P.Maris,$Phys$.$Rev$.{\bf{D54}} (1996) 4049. \\
\\
$[8]$  K.-I.Kondo,preprint OUTP-96-50P,CHIBA-EP-96;hep-th/9608402. \\
\\
$[9]$ N.Dorey and N.E.Mavromatos,$Phys$.$Lett$.{\bf{B266}} (1991) 163.\\
\\
$[10]$ I.J.R Aitchison, N.Dorey, M.Klein-Kreisler and N.E Mavromatos,
$Phys$. $Lett$. {\bf{B294}} (1992) 91. \\
\\
$[11]$ I.J.R Aitchison and M.Klein-Kreisler, $Phys$. $Rev$. {\bf{D50}} (1993) 
 1068.\\
\\
$[12]$ I.J.R Aitchison,$Z.Phys.$ {\bf{C67}} (1995) 303.\\
\\
$[13]$ N.Dorey and N.E Mavromatos, $Nucl$. $Phys$. {\bf{B386}} (1992) 614.\\
\\
$[14]$ A.Kovner and B.Rosenstein, $Phys$. $Rev$. {\bf{B42}} (1990) 4748.\\
A.Dorey and N.E Mavromatos, $Phys$. $Lett$. {\bf{B250}} (1990) 107.\\
G.W Semenoff and N.Weiss, $Phys$. $Lett$. {\bf{B250}} (1990) 117.\\

\newpage

\noindent {\bf Figure Captions}\\

Figure 1. Contributions to $\Delta_{\mu \nu}$ to leading order in $1/N$\\

Figure 2 (a).  $\Pi_1$ as a function of explicit $m$ and 3-momenta,$p_{b}$.\\

Figure 2 (b). $\Pi_2$ as a function of explicit $m$ and 3-momenta,$p_{b}$.\\

Figure 2 (c). $\Pi_3$ as a function of explicit $m$ and 3-momenta,$p_{b}$\\

Figure 3 (a). Top graph: The numerical data points and the approximation $\beta p_b/8$ for the $m \neq 0$ modes of $\Pi_1(m=1,p_b)$,shown for $p_b \beta$ ranging from $0-4$. Bottom graph: The numerical data points and the approximation $\beta p_b/8$ for the $m \neq 0$ modes of $\Pi_1(m=1,p_b)$,shown for  $p_b \beta$ ranging from $0-36$.\\

Figure 3 (b). Top graph: The numerical data points and the approximation $\beta p_b/8$ for the $m \neq 0$ modes of $\Pi_3(m=1,p_b)$, shown for $p_b \beta$ ranging from $0-9$. Bottom graph: The numerical data points and the approximation $\beta p_b/8$ for the $m \neq 0$ modes of $\Pi_3(m=1,p_b)$, shown for  $p_b \beta$ ranging from $0-36$.\\

Figure 4. The numerical results for the constant mass calculation as functions of $T \over k_B \alpha$ for $N=0.5,1,1.5$.\\

Figure 5. Plot of $k_BT_c / \alpha$ vs. $N$ for the constant mass calculation; the points are the numerical values calculated from (21),the curve is an approximate form for $T_c(N)$ as given by equation (28).\\

Figure 6. The numerical results for the frequency dependent solution, $\Sigma(p_{0f},T)$ as a function of $p_{0f}$ for various $k_BT/\alpha$ at $N=1$.\\

Figure 7. A comparison of $\Sigma_{const}$, the mass for the constant mass calculation ,with $\Sigma_{freq}(m=0,T=0)$, the zero-frequency 
mass for the frequency dependent calculation, as functions of $N$.\\

Figure 8 (a).The numerical results for the frequency dependent mass calculation as functions of $T \over k_B \alpha$ for $N=0.5,0.7,1$.\\

Figure 8 (b).The numerical results for the frequency dependent mass calculation as functions of $N$ for $Tk_B/\alpha=0,0.001,0.002,0.004$, used to estimate $N_c$ in the case of $Tk_B/\alpha=0$.\\

Figure 9. A phase diagram for the frequency dependent solution, showing the boundary between the massless and massive fermion phases in the $N-T$ plane.\\

\noindent {\bf Table Captions}\\ 

Table 1. The values of $r$ calculated for various $N$ in the frequency dependent mass calculation.\\

\newpage

\begin{figure}
\begin{center}
\includegraphics{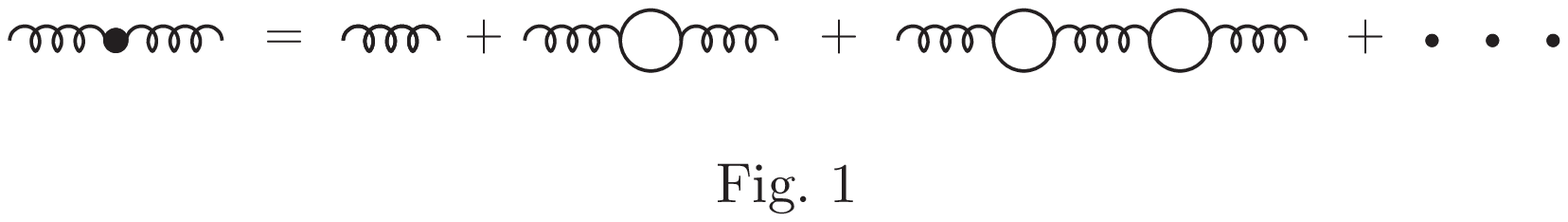}
\end{center}
\end{figure}

\begin{figure}
\begin{center}
\includegraphics{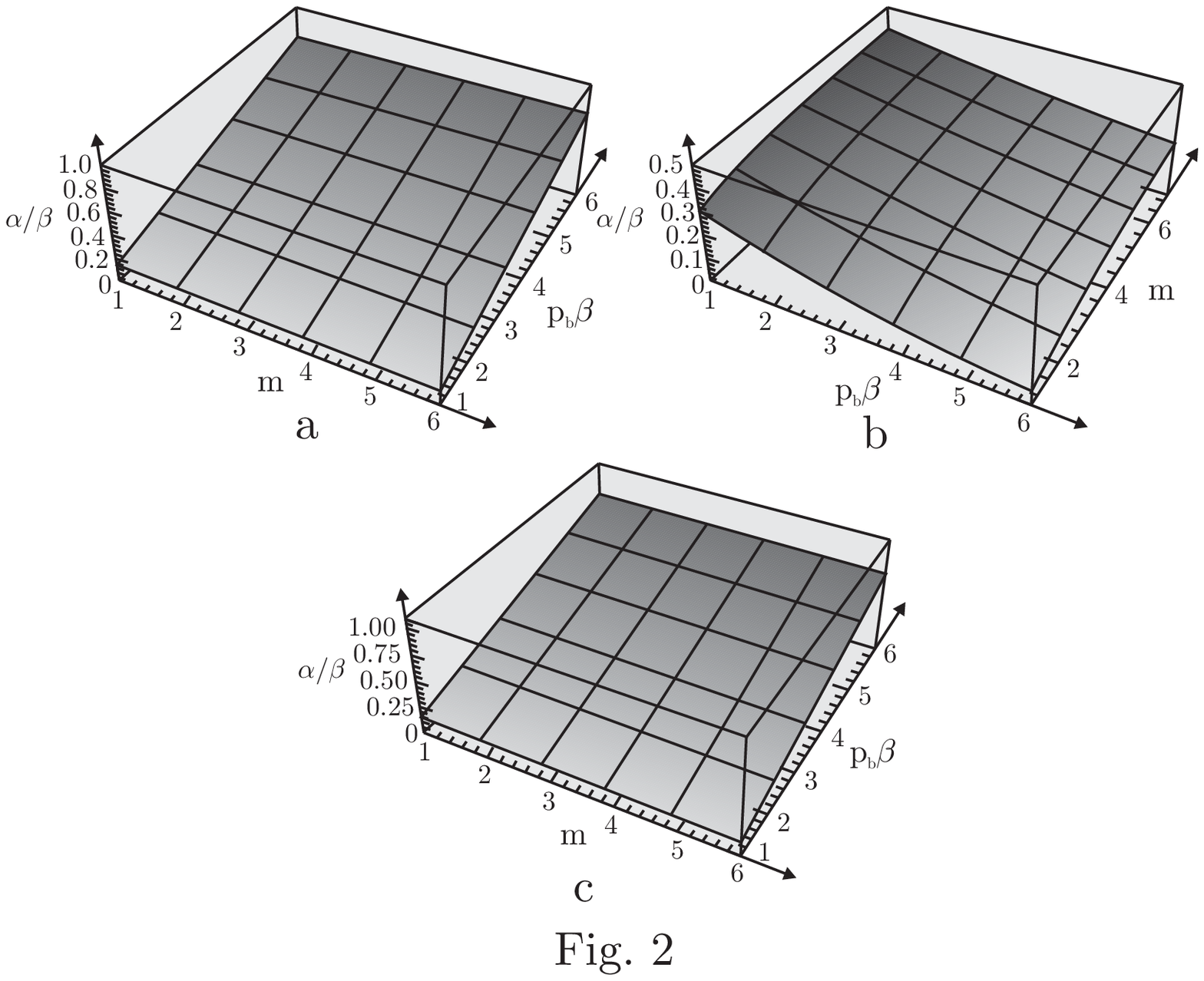}
\end{center}
\end{figure}

\begin{figure}
\begin{center}
\includegraphics{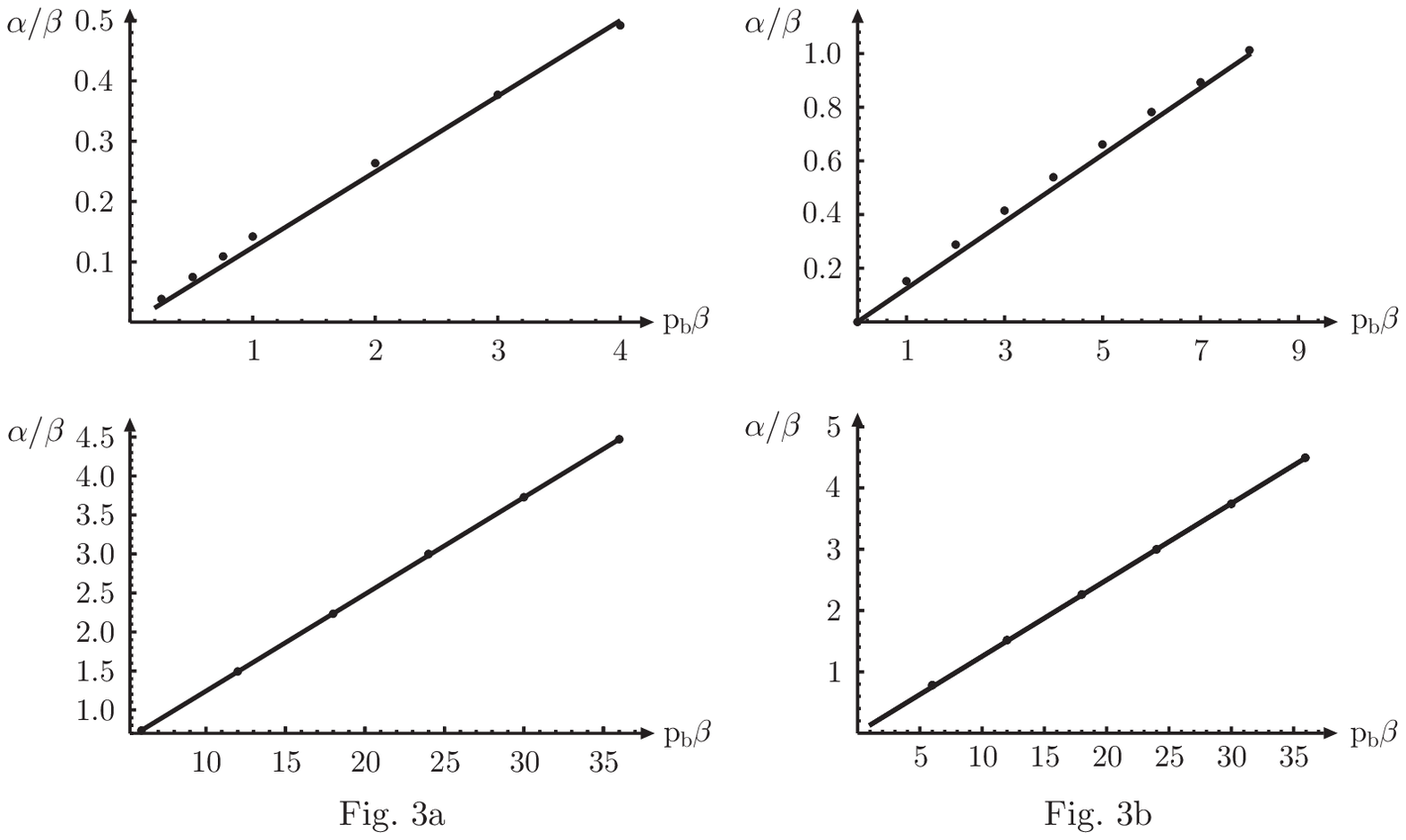}
\end{center}
\end{figure}

\begin{figure}
\begin{center}
\includegraphics{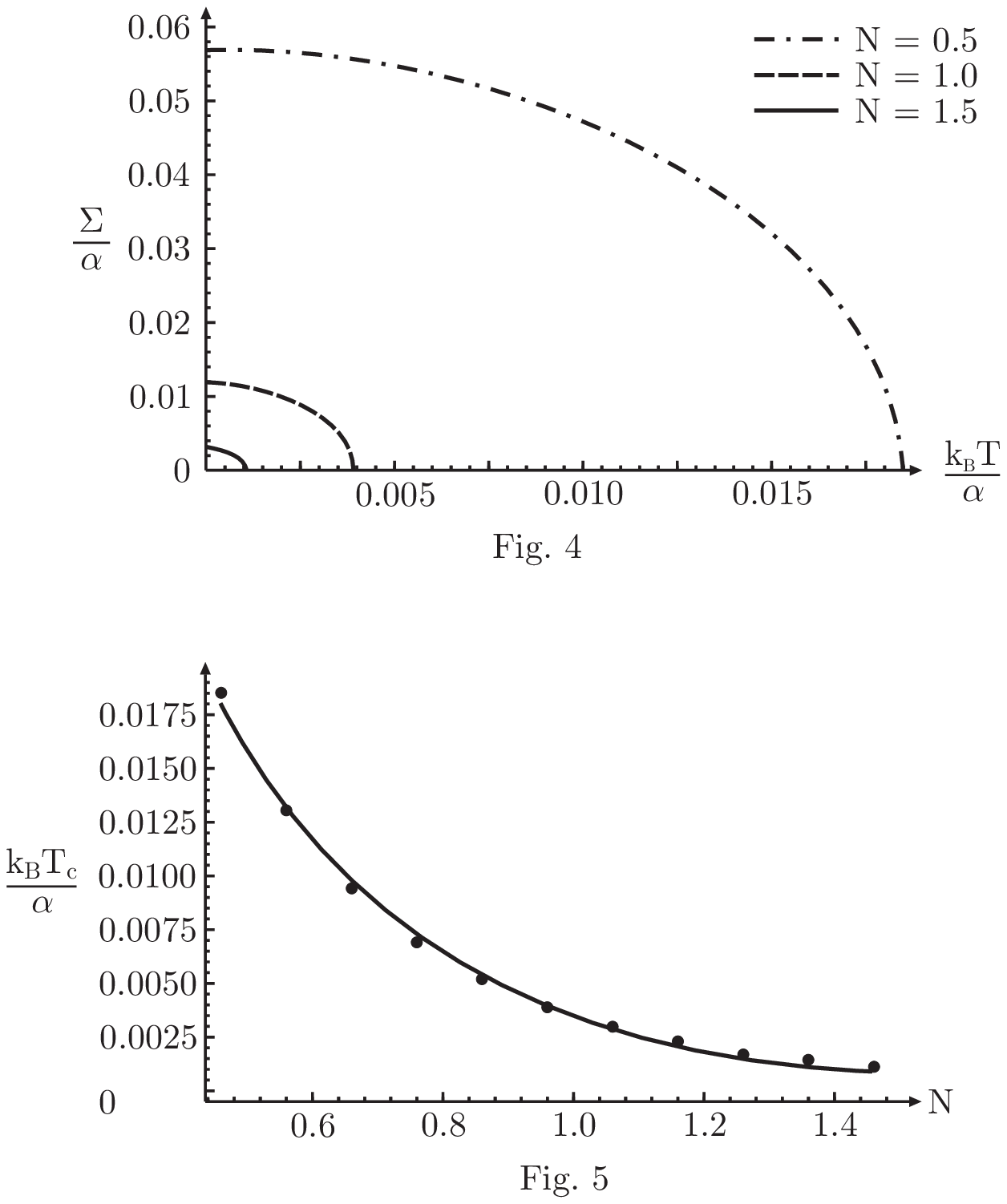}
\end{center}
\end{figure}

\begin{figure}
\begin{center}
\includegraphics{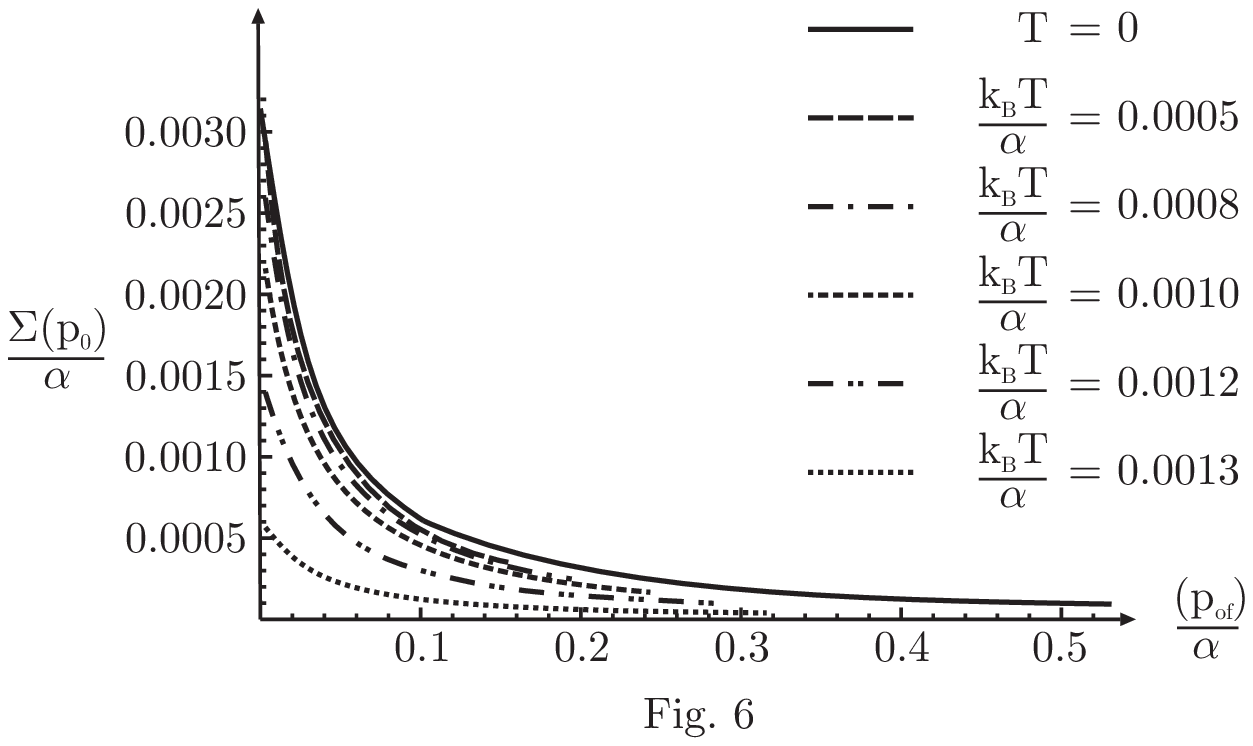}
\end{center}
\end{figure}

\begin{figure}
\begin{center}
\includegraphics{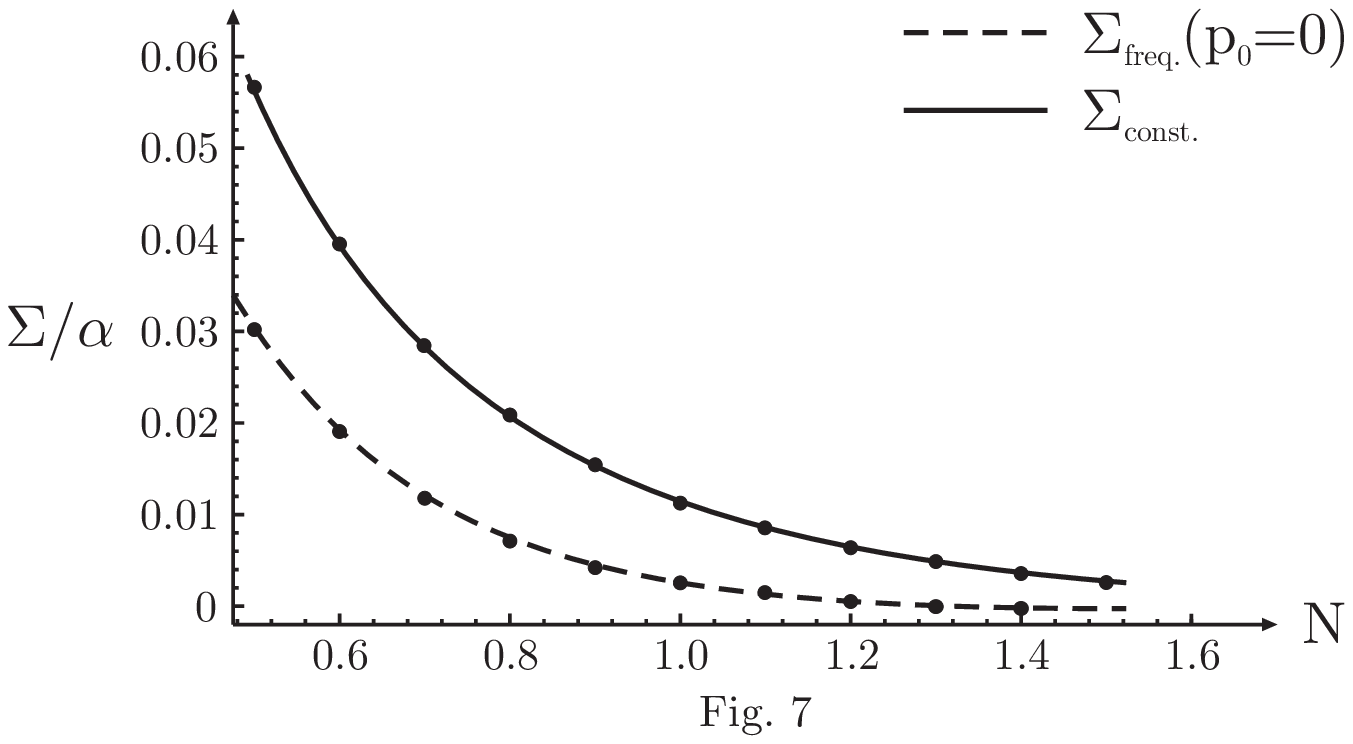}
\end{center}
\end{figure}

\begin{figure}
\begin{center}
\includegraphics{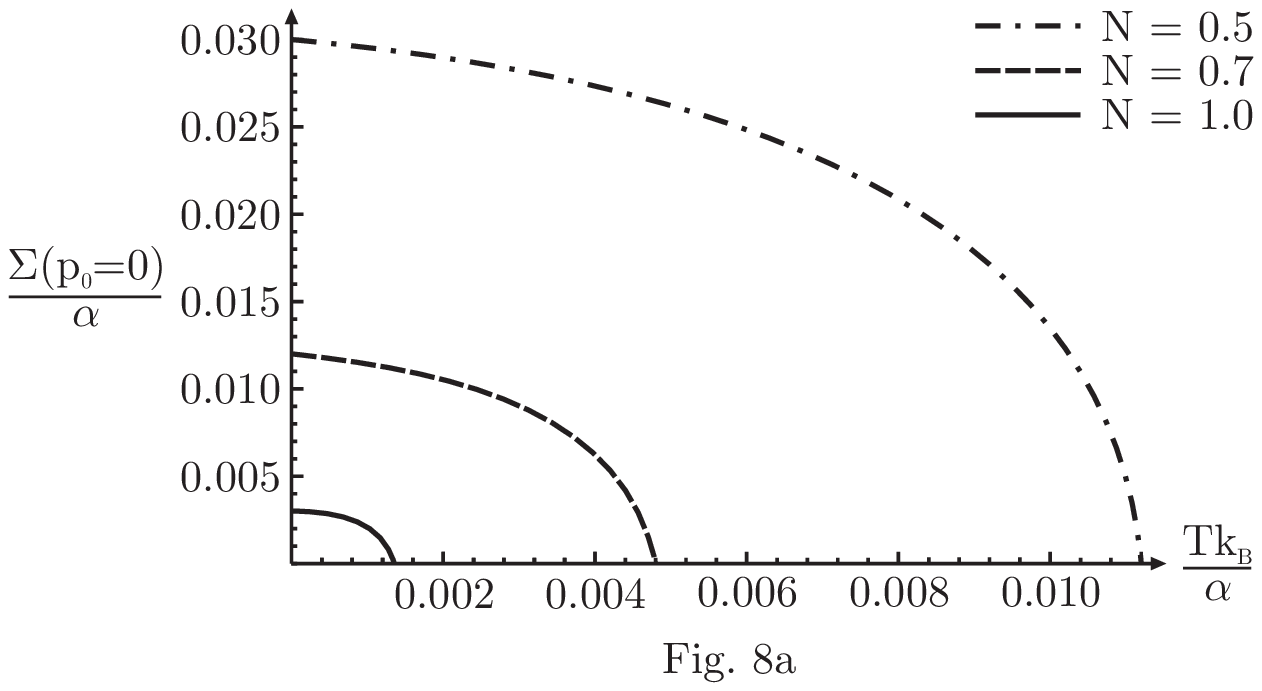}
\end{center}
\end{figure}

\begin{figure}
\begin{center}
\includegraphics{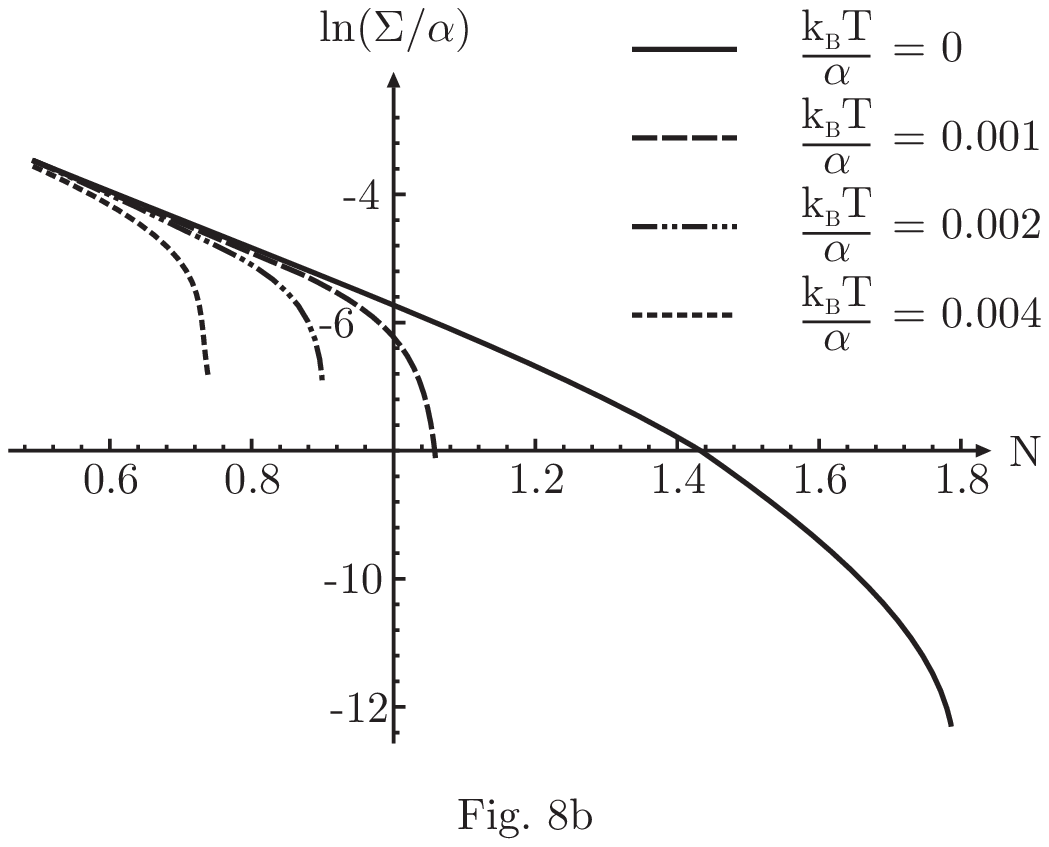}
\end{center}
\end{figure}

\begin{figure}
\begin{center}
\includegraphics{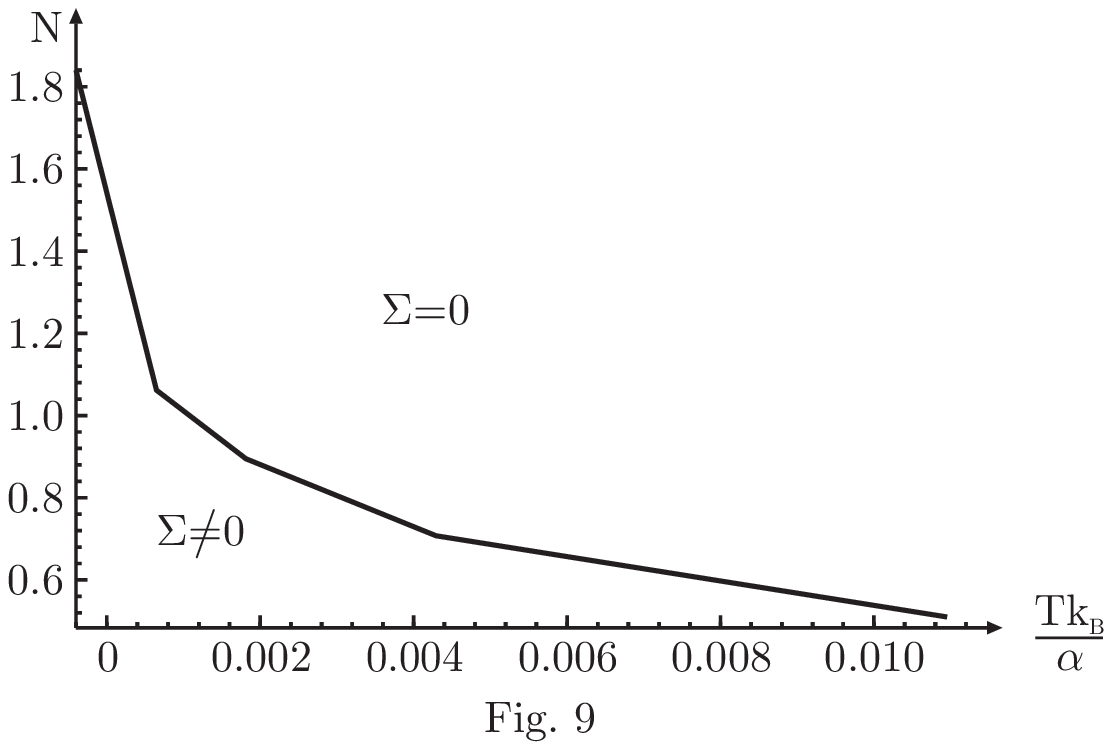}
\end{center}
\end{figure}

\end{document}